\documentclass[10pt,twocolumn,letterpaper]{article}
\usepackage{wacv}
\usepackage{times}
\usepackage{epsfig}
\usepackage{graphicx}
\usepackage{amsmath}
\usepackage{amssymb}
\usepackage{float}
\usepackage{tabularx}
\usepackage{bbm}
\usepackage{breqn}
\usepackage{booktabs}
\usepackage{authblk}
\usepackage{caption}
\usepackage[pagebackref=true,breaklinks=true,letterpaper=true,colorlinks,bookmarks=false]{hyperref}

\wacvfinalcopy

\def\wacvPaperID{****} 

\begin{document}

\title{Atrial Fibrillation Detection Using RR-Intervals for Application in Photoplethysmographs}
\author[1]{Georgia Smith}
\author[2]{Dr. Yishi Wang}
\affil[1]{The University of Virginia's College at Wise}
\affil[2]{The University of North Carolina at Wilmington}

\vspace{-9cm}
\maketitle
\vspace{-1cm}
\begin{abstract}
Atrial Fibrillation is a common form of irregular heart rhythm that can be very dangerous. Our primary goal is to analyze Atrial Fibrillation data within ECGs to develop a model based only on RR-Intervals, or the length between heart-beats, to create a real time classification model for Atrial Fibrillation to be implemented in common heart-rate monitors on the market today. Physionet's MIT-BIH Atrial Fibrillation Database \cite{goldberger2000physiobank} and 2017 Challenge Database \cite{clifford2017af} were used to identify patterns of Atrial Fibrillation and test classification models on. These two datasets are very different. The MIT-BIH database contains long samples taken with a medical grade device, which is not useful for simulating a consumer device, but is useful for Atrial Fibrillation pattern detection. The 2017 Challenge database includes short ($<60sec$) samples taken with a portable device and reveals many of the challenges of Atrial Fibrillation classification in a real-time device. We developed multiple SVM models with three sets of extracted features as predictor variables which gave us moderately high accuracies with low computational intensity. With robust filtering techniques already applied in many Photoplethysmograph-based consumer heart-rate monitors, this method can be used to develop a reliable real time model for Atrial Fibrillation detection in consumer-grade heart-rate monitors.
\vspace{.3cm} \\
    \emph{Keywords:} Atrial Fibrillation Classification, Support Vector Machine, Feature Extraction, Real Time Model, Physionet, Electrocardiogram, Photoplethysmogram, MIT-BIH, Physionet, Consumer Implementation, RR-Intervals, Transition Matrix
\end{abstract}


\section{Introduction}
Atrial Fibrillation (AFib) is the most common form of Arrhythmia, or irregular heart rhythm. It can be very dangerous, causing an increased risk of stroke and heart attack/failure \cite{nihbli}. Symptoms of Atrial Fibrillation include increased heart rate, heart palpitations, and low blood pressure. Atrial Fibrillation is characterized by two main features: an absence of P-waves in an Electrocardiogram (ECG) Recording, and “Irregular Irregularity,” meaning that there is no pattern to when AFib rhythms start or stop and no specific pattern to the heart rhythm. The irregularity between heartbeat in Atrial Fibrillation is caused by abnormal electrical activity in the atrial, or top, chambers of the heart.

\includegraphics[width = .45\textwidth, keepaspectratio]{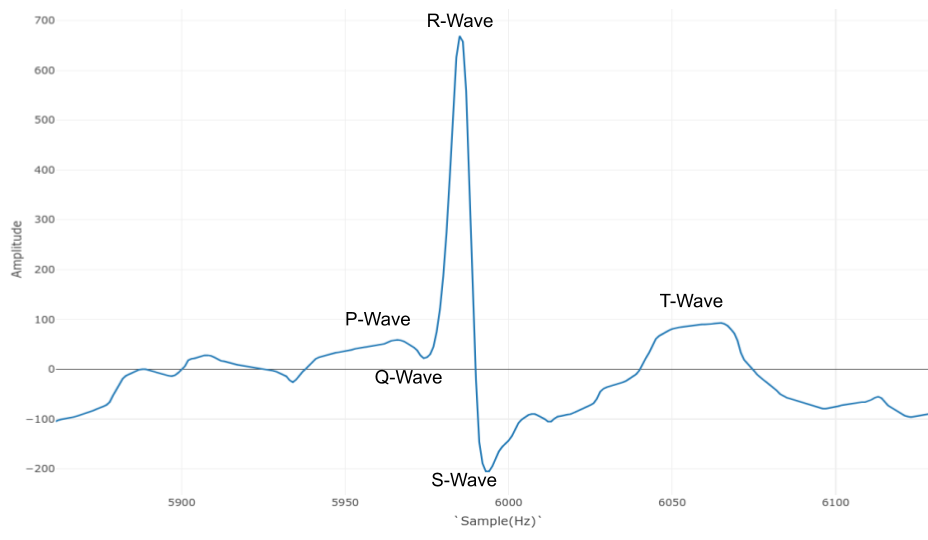}

The P-Wave, which characterizes Atrial Fibrillation, has a low amplitude in comparison to the R-Wave which is the peak of ECG recordings. We chose to use the interval between adjacent R-wave peaks (RR-Interval) to detect Atrial Fibrillation because P-Waves are often misidentified due to noise within Electrocardiogram recordings and because in Photoplethysmographs (PPGs), which are used in consumer heart-rate monitors, just the heart beat is identified. To analyze and identify RR-Interval patterns of ECG recordings of individuals experiencing AFib versus Normal heart-beat rhythms we used the MIT-BIH Atrial Fibrillation Database  \cite{goldberger2000physiobank} and the Physiobank 2017 Challenge Dataset \cite{clifford2017af}. The MIT-BIH Database includes 230 hours of Electrocardiogram (ECG) recordings with Atrial Fibrillation labelled, which was used to learn about AFib patterns and test potential models on. The MIT-BIH Database has been used by many authors to detect Atrial Fibrillation due to its robust and longitudinal nature \cite{young1999comparative, moody1983, lake2010accurate, rubio, ghodrati2008statistical}. The MIT-BIH Database has also been thoroughly cleaned for accurate model building. The 2017 Challenge Dataset is less thoroughly cleaned set of recordings which allowed us to test the robustness of models.

We first reviewed and considered other research on Atrial Fibrillation classification from different academics. We then focused on initial analysis of the data where we extracted and cleaned the data from the MIT-BIH Atrial Fibrillation Database and the 2017 Challenge Dataset, identifying patterns of AFib in the ECG recordings, and developing potential modelling techniques to detect dangerous levels of Atrial Fibrillation. Once the data was extracted we developed multiple features and then subset the data into 45-second intervals, including the extracted features. We then used these samples to test classification models.

\subsection{Goal}

Atrial Fibrillation is difficult to diagnose because of its irregular nature. Often diagnosis of Atrial Fibrillation is incidental, which is not ideal given AFib can increase the risk of stroke in adults older than 45 by 20\% \cite{rubio}. Our goal is to address this by developing a simple, accurate model for real time Atrial Fibrillation detection using common heart-rate monitors on the market today (Apple Watch, FitBit, etc.). These aforementioned heart-rate monitors often do not use Electrocardiograms, but instead use Photoplethysmograms (PPGs), which measure the expansion and contraction of blood vessels as the heart pumps blood throughout the body. This technology is cheaper and easier for day to day use because it can be placed anywhere that there is a blood vessel. While application in PPGs is the goal, we are using ECG data to identify patterns of AFib through RR-intervals to simulate PPG signals because PPG data from Atrial Fibrillation patients is not publicly accessible.

\subsection{Contribution}

\begin{itemize}
    \item Cross-database Transfer Learning \vspace{-.2cm}
    \item Transition Matrix Predictor Variables \vspace{-.2cm}
    \item $RR_{100}$ modification to $RR_{200}$ \vspace{-.2cm}
    \item Outlier Removal \vspace{-.2cm}
    \item $RR_{var}$ modifications \vspace{-.2cm}
    \item Implementation for Electrocardiograms and Photoplethysmographs \vspace{-.2cm}
    \item RR-Interval-based Predictor Variables \vspace{-.2cm}
\end{itemize}
\section{Literature Review}
Electrocardiograms provide immense amount of data to work with in identifying Atrial Fibrillation and, with Physiobank's open source Electrocardiogram databases, research into automated classification of Atrial Fibrillation and other heart conditions is very thorough. While there is much literature out there for Atrial Fibrillation, our main focus is on extraction of features based on RR-intervals with low computational intensity for implementation in a real time PPG/ECG device.

\subsection{Moody and Mark}

Moody and Mark, in their 1983 paper, attempted to predict future heart-beat rhythms given previous heart-beats in order to identify patterns of Atrial Fibrillation. They did this by creating a running mean and classifying the individual RR-Intervals as long, normal, or short by comparison to this running mean \cite{moody1983}.

\vspace{-.3cm}
            \begin{equation}
        rmean_i = 0.75*rmean_{i-1} + 0.25*RR_i
        \label{moddyrmean},
            \end{equation}
\vspace{-.3cm}
            \begin{equation}
    C_i = 
        \begin{cases}
        short & \text{if  } RR_i < 0.85 \times rrmean_i \\
        long & \text{if  } RR_i > 1.15\times rrmean_i \\
        normal & \text{otherwise}
        \end{cases}
        \label{moodyclass}.
            \end{equation}

 They then used the classifications to develop transition matrices for a Markov model to predict the next heart beat rhythm. While the prediction of the next rhythm is not our concern, the transition matrices Moody and Mark utilized allow a quantitative measurement for how much the heart-beat varies from a ``normal" heart rhythm.

 \subsection{Rubio and Berega\~na}

In their 2011 paper, Rubio and Berega\~na \cite{rubio} compared 9 different extraction and classification techniques used for Atrial Fibrillation classification. First they filtered their ECG data using a high pass filter to remove Baseline Wander, a common noise in ECG data due to patient movement and incorrect ECG node placement. They then set a maximum and minimum threshold for reading R-Peaks to reduce incorrect R-Peak detection due to noise. The cleaning methods were very simple, yet allowed for more accurate readings even in the presence of noise. The simplicity of these cleaning methods could be very useful in implementation of a real-time AFib-Detection device. Once Rubio and Berega\~na cleaned the ECG data they tested 9 extraction techniques. Out of the 9 extracted measurements tested we identified two that may be useful: Ghodrati et al \cite{ghodrati2008statistical} and Young et al \cite{young1999comparative}. 

Ghodrati et al used a similar algorithm to Moody and Mark, except instead of classifying the RR-Intervals into transitions matrices they developed an RR-variance algorithm defined by

            \begin{equation}
                RRvar_i = \frac{|RR_i - RR_{i-1}|}{rmean_i}
                \label{rrvar},
            \end{equation}
            and $\overline{RR_l}$ is defined recursively by 
            \begin{equation}
                rmean_i = 0.9*\overline{RR_{i-1}}+0.1*RR_i
                \label{barrr}.
            \end{equation}

Instead of quantifying how much a heart-beat varies from ``normal" with transitions, Ghodrati et al's $RRvar$ gives a longitudinal measurement similar to a running mean which quantifies how much RR-Intervals are changing. 

Young et al \cite{young1999comparative} used a much simpler classification technique using no statistical models, but instead a simple classification algorithm using counting defined as the $RR_{100}$ algorithm.

            \begin{align}
                \label{countit}
                \begin{split}
                RR_i - RR_{i-1} > 100ms \rightarrow count = count+1
                , \\
                RR_i - RR_{i-1} < 100ms \rightarrow count = count-1 ,
                \end{split}
            \end{align}
            \begin{equation}
               count > 6 \rightarrow AFib
               \label{countclass}.
            \end{equation}

We identified the $count$ variable, in Equations \ref{countit} and \ref{countclass}, as a potential covariate for our model. 

\section{Data}

\subsection{The MIT-BIH Database}

The MIT-BIH Atrial Fibrillation Database includes 23, complete, 10-Hour ECG Recordings from individuals diagnosed with Atrial Fibrillation. The dataset is oversampled with AFib data, which is useful in building a model, but not for an accurate testing set. The dataset includes four rhythm types, Atrial Fibrillation (AFib), Normal (N), Atrial Flutter (AFL), and AV Junctional Rhythm (J).

\begin{table}[ht]
    \centering
    \begin{tabular}{lcccc}
        \toprule
         & \textbf{AF}ib & \textbf{AFL} & \textbf{J} & \textbf{N} \\
         \midrule
        \textbf{Proportion of Data} & 0.398 & 0.006 & 0.004 & 0.592 \\
        \bottomrule
    \end{tabular}
    \vspace{.2cm}
    \caption{Proportion of Different Heart-Rhythms in MIT-BIH Data}
    \label{tab:mitbihrhyth}
\end{table}

The recordings include RR-Intervals, annotations sectioning heart-beats into categories, and digitized signals for each subject. The readings were taken at 250 samples per second to increase their reliability. On average, the recordings include 48,000 heart-beats. The data in the MIT-BIH Atrial Fibrillation Database is subset into three files: reference annotation files, beat annotation files, and digitized signal files. For this analysis we focused on the reference annotation files (.atr files) and the beat annotation files (.qrs files) to extract RR-Intervals and annotations. 

The MIT-BIH Database is a very reliable Atrial Fibrillation Database that is often used in the field of Atrial Fibrillation and other heart-rhythm based research \cite{rubio, lake2010accurate, moody1983}. This dataset is so reliable because it have been cleaned and tested in multiple research studies. Physionet also used two or more medical professionals to label the rhythms for every ECG recording.

\subsection{Physionet 2017 Challenge Data}

In 2017 Physionet, the same organization which hosts the MIT-BIH Database, held a competition for Atrial Fibrillation Classification. A main focus of this competition was to find methods of Atrial Fibrillation Detection without over-sampled, clean data and develop methods to identify more than two classes. Therefore, the data in the Physiobant 2017 Challenge Dataset includes four classifications: Noise, Atrial Fibrillation, Normal, and Other.

\begin{table}[ht]
    \centering
    \begin{tabular}{lccccc}
        \toprule
        & \textbf{N}ormal & \textbf{AF} & \textbf{O}ther & \textbf{N}oisy & \textbf{T}otal\\
        \midrule
        \textbf{Amount} & 5,154 & 771 & 2,557 & 46 & 8,528 \\
        \bottomrule
    \end{tabular}
    \vspace{.2cm}
    \caption{Types of Rhythm within 2017 Competition Data}
    \label{train2017}
\end{table}

The Noisy category includes all ECG recordings which are unreadable due to an excess of noise and the Other category includes all identifiable heart-rhythms that are not AFib or Normal.The 2017 Challenge dataset also has a range of times from 9-60 seconds, although most recordings are 30 seconds long. The varying times cause our samples to be varying in length causing an extra variable that we most likely would not have to account for in the continuous model/device which we hope to develop.

\begin{figure}[H]
    \centering
    \includegraphics[width = .48\textwidth, keepaspectratio]{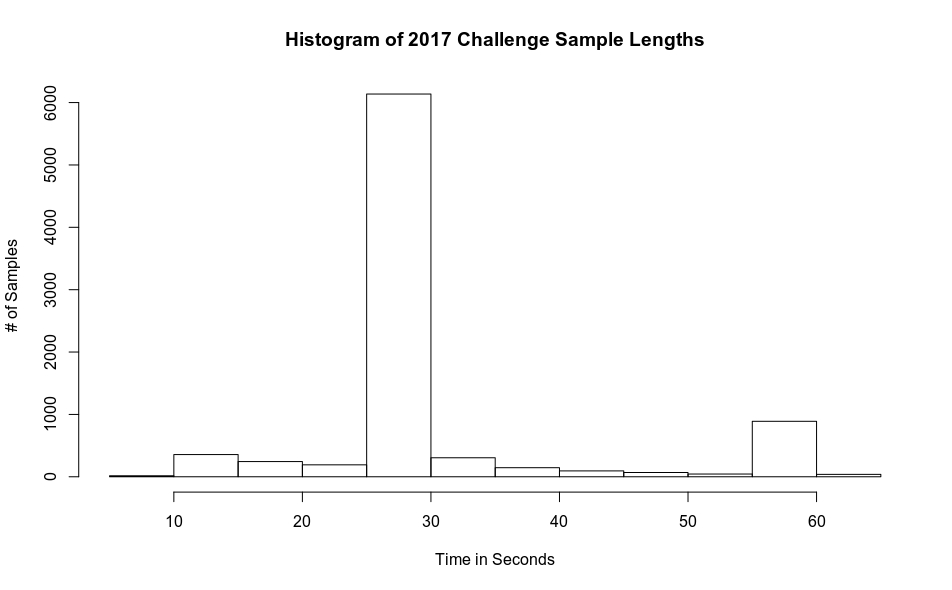}
    \caption{Distribution of Sample-Lengths in 2017 Challenge Dataset}
    \label{fig:hist2017samp}
\end{figure}

In the 2017 Challenge dataset the recordings were taken with a single lead, portable Electrocardiogram device developed by AliveCor, causing further problems of R-Peak detection due to noise. Because this data was taken with a portable device,  it is a more accurate test set for future implementation of AFib classification models.

\section{Data Preprocessing and Feature Extraction}

\subsection{Extraction of RR-Intervals}

Both the MIT-BIH Database and the 2017 Challenge Dataset came in digitized formats for use with Moody's Wave-Form-Database (WFDB) \cite{wfdb}. To extract the RR-Intervals from these files we also used the WFDB of Functions and extraction methods. On the 2017 Challenge Dataset we first had to extract .dat files from the digitized matlab files provided using ``mat2wfdb." We then had to extract .qrs files from the .dat files for the 2017 Challenge Dataset using WFDB's ``wqrs" function. We then had both the MIT-BIH dataset and the 2017 Challenge dataset in .qrs form where RR-Intervals could be detected. We first used the ``ann2rr," or annotations to RR-Intervals, function to extract RR-Intervals from the .qrs files. The data extracted from this function included the sample where each heart-beat started and ended, as well as the RR-Intervals in seconds. Then, to extract the labelled heart-beat rhythms from the MIT-BIH dataset we used the ``rdann," or read annotations, function to extract heart-beat annotations for the .atr files. The rhythms for the 2017 Challenge dataset were labelled in a separate .csv provided by Physionet. The 2017 Challenge data heart-beat rhythms were classified individually by classifying the entire sample as a single rhythm due to the short recording times.

The data extracted from ``rdann" included elapsed time at start, sample number at start, and annotation. The annotations classified the MIT-BIH heart-beats into four rhythm categories: Atrial Fibrillation (AFib), Normal (N), Atrial Flutter (AFL), and AV Junctional Rhythm (J). Because the annotations were formatted by range they had to be merged the RR-Intervals from the 23 MIT-BIH datasets. To merge the files we used the sample numbers in the extracted files to set each group of RR-Intervals as their corresponding annotation.

\subsection{Signal Processing}

When looking at the 2017 Challenge dataset an abnormal amount of noise was present in comparison to the MIT-BIH database. When extracted the same way as the MIT-BIH dataset the WFDB \cite{wfdb} extracted less than 8 RR-Intervals for samples with 30 or more seconds of data due to noise. The noise present is accounted for because the 2017 Challenge dataset was collected using a portable, single-lead ECG. The MIT-BIH database was collected using a medical-grade, dual-lead ECG in a constant setting for accurate recordings. Taking this into consideration we looked into signal processing methods for the 2017 Challenge dataset.

\begin{figure}[H]
    \centering
    \includegraphics[width = .47\textwidth, keepaspectratio]{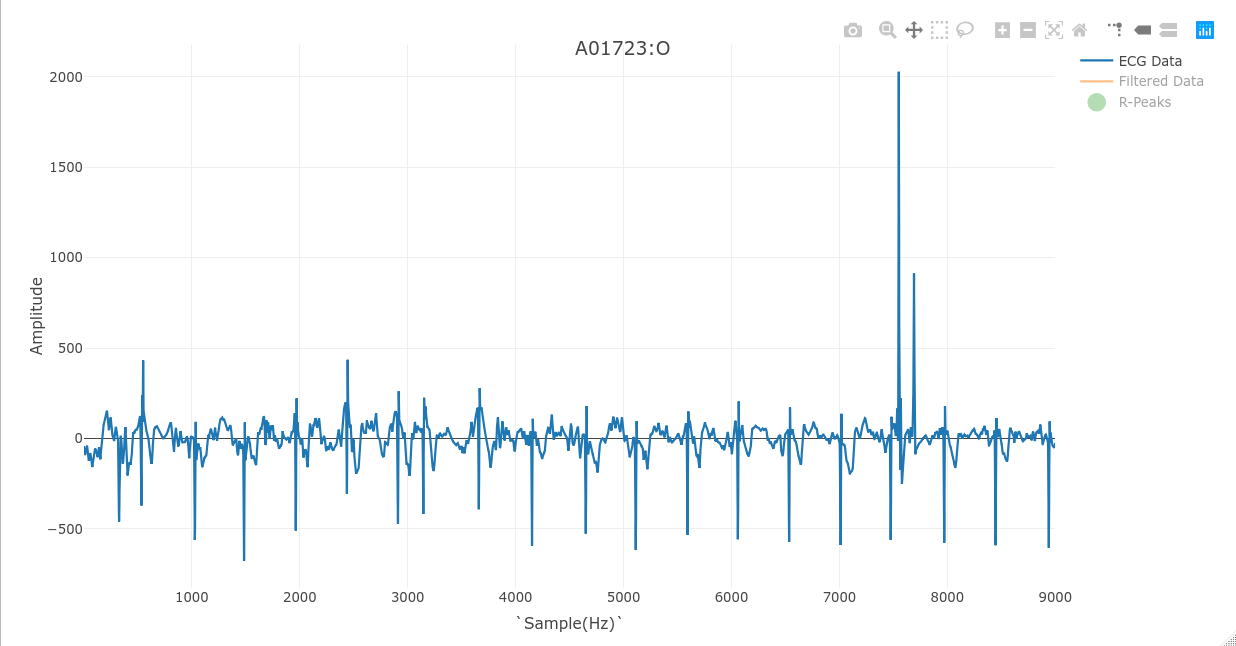}
    \caption{Subject A01723 Raw Data, with $<8$ RR-Intervals observed by WFDB, Heart Rhythm Classified as Other}
    \label{fig:filternone}
\end{figure}

In 2011, Rubio and Berega\~na used high pass filters on the MIT-BIH database for their testing of Atrial Fibrillation Classification Models to reduce noise and baseline wander. Taking this into consideration we applied low pass and high pass filtering techniques used in unison as a bandpass filter on the 2017 Challenge Dataset. 

\begin{figure}[H]
    \centering
    \includegraphics[width = .47\textwidth, keepaspectratio]{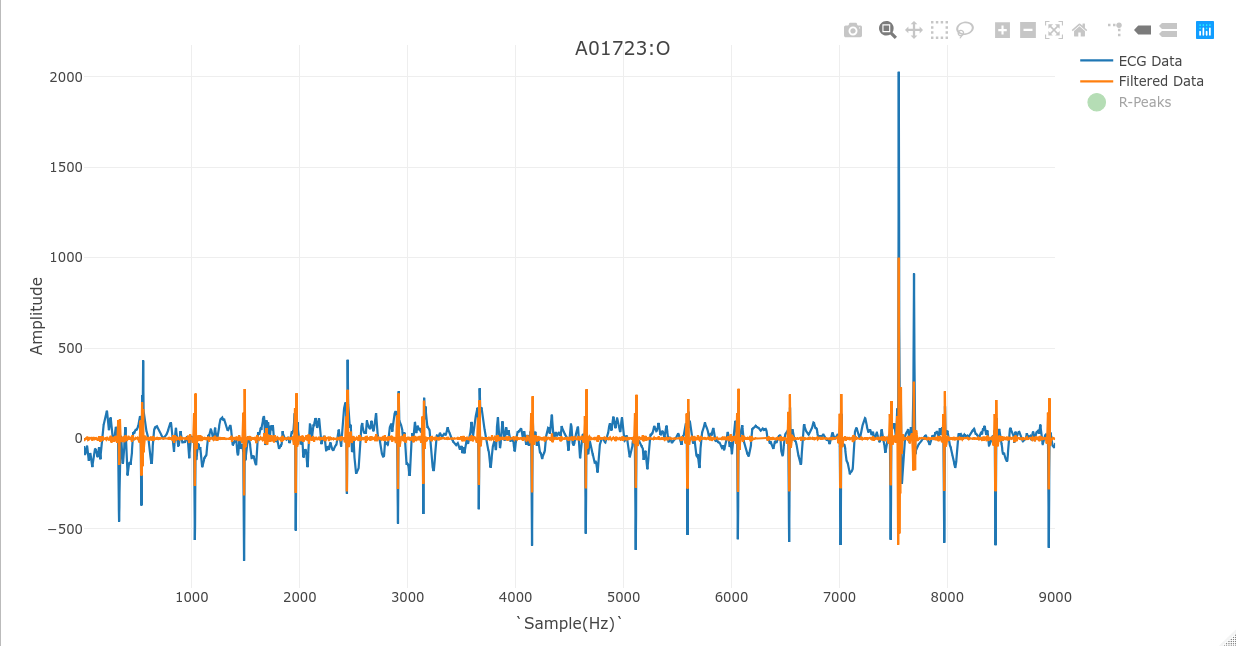}
    \caption{Subject A01723 Raw Data Overlayed by Filtered Data}
    \label{fig:filternorr}
\end{figure}

This resulted the noise to be decreased with only R-Peaks and other extreme noise readings to be peaks on the ECG. 

\begin{figure}[H]
    \centering
    \includegraphics[width = .47\textwidth, keepaspectratio]{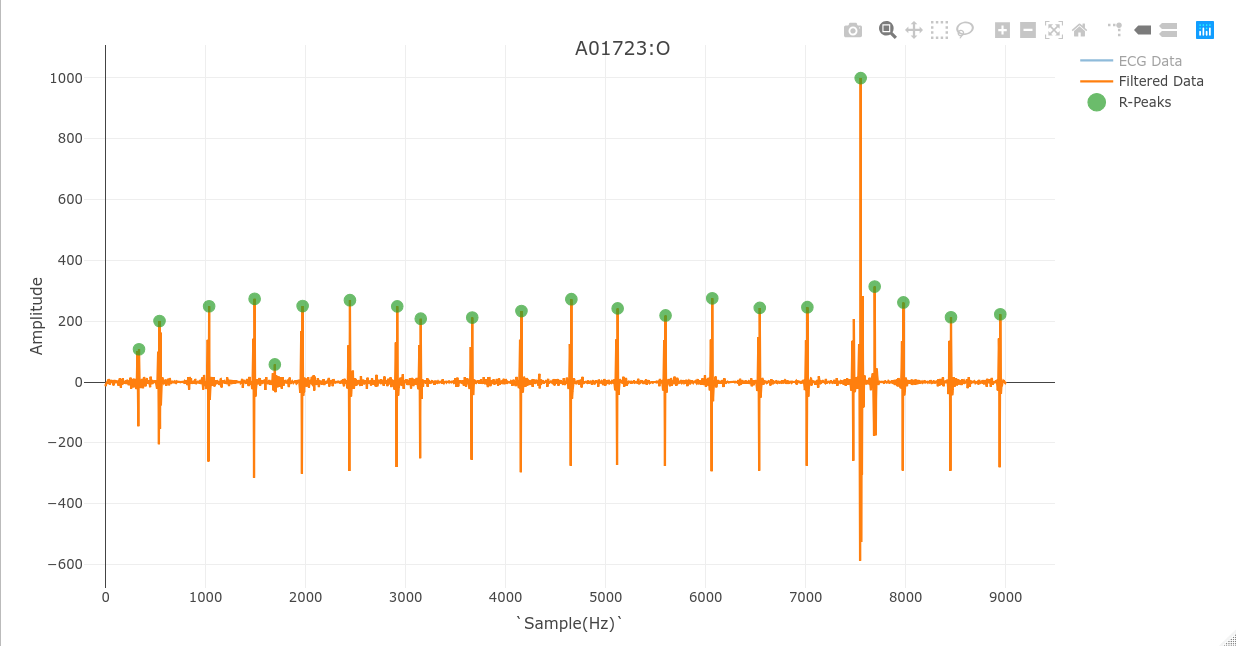}
    \caption{Subject A01723 Filtered Data with R-Peaks Detected}
    \label{fig:giltRR}
\end{figure}

This did not remove all problems though, ECGs with noise at an extreme amplitude still resulted in incorrect R-Interval readings. To address this we developed outlier thresholds at $\text{heart-rate}>200bpm$, $\text{heart-rate}<20bpm$, and $\overline{drr}>1sec$, where $\overline{drr}$ is the arithmetic mean of the difference in RR-Interval length defined as
\begin{equation}
    \overline{drr} = \frac{|RR_i - RR_{i-1}|}{\Sigma RR}.
    \label{drr}
\end{equation}

Any sample from the 2017 Challenge Dataset which was classified by one or more of the thresholds was deemed an outlier.

\begin{figure}[H]
    \centering
    \includegraphics[width = .47\textwidth, keepaspectratio]{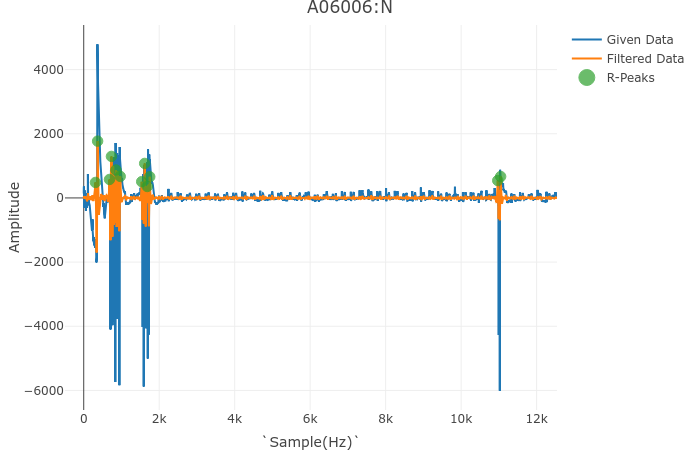}
    \caption{Outlier Example, Subject A06006, Heart-Rhythm Classified as Normal}
    \label{fig:out}
\end{figure}

With these outliers specified we were able to further filter just those recordings that had noise distorting R-Peak identifications by removing sections with abnormal noise, which were usually isolated at the beginning and end of every recording. Once we removed those sections we processed the remaining data with a bandpass filter and identified R-Peaks. We then checked for outliers once more and found the data to be more reliable in detecting R-Peaks.

\begin{figure}[H]
    \centering
    \includegraphics[width = .47\textwidth, keepaspectratio]{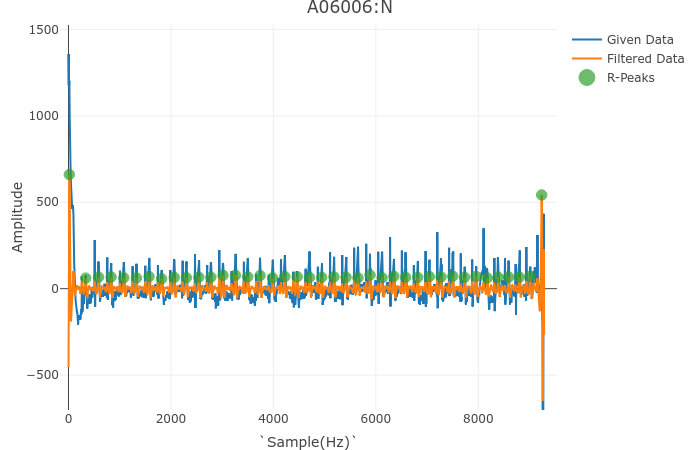}
    \caption{Subject A06006 R-Peak Detection with Noise Removed}
    \label{fig:cleanout}
\end{figure}

\subsection{Subsetting}

Once the RR-Intervals were extracted we removed any RR-Intervals which were more than four standard deviations away from the mean. This was done because there are recordings within the MIT-BIH dataset which have bad readings marked as no heart-beat, resulting in some RR-Intervals being 5-10 seconds. With outliers removed from both datasets, we applied Moody and Mark's running mean for each RR-Interval defined recursively by Equation \ref{moddyrmean}. A running mean was used because of the longitudinal nature of heart-beat data and the objective of this analysis to develop a method for real-time AFib detection. This running mean was also determined to increase accuracy by Moody and Mark. We then classified each individual RR-Interval as short, normal, or long based defined by Equation \ref{moodyclass}. This process done on the MIT-BIH dataset and the 2017 Challenge dataset.

With the RR-Interval Classifications we then developed transition matrices for each subject. The transition matrices tallied each transition (Short-to-Short, Short-to-Normal, Short-to-Long, etc.) for future reference. 

The complete, merged, data for each subject now included start and stop samples, RR-Interval in seconds, Annotations (within the  MIT-BIH Dataset), running mean, and classification as short, normal, or long. For the MIT-BIH Database we subset each ECG recording in 45-second samples for better simulation of real-time classification. There was no need to subset the 2017 Challenge dataset due to the already short sample lengths. Within each subset of the MIT-BIH data and recording from the 2017 Challenge data we developed a transition matrix based on the RR-Interval classification.

\begin{table}[ht]
    \centering
    \begin{tabular}{llccc}
        & && \textbf{From}&\\
    \toprule
         && \textbf{S}hort & \textbf{N}ormal & \textbf{L}ong \\
         \midrule
         &\textbf{Short} &  8 & 6 & 7 \\
         \textbf{To}&\textbf{Normal} & 9 & 14 & 6 \\
         &\textbf{Long} & 4 & 10 & 4 \\
         \bottomrule
    \end{tabular}
    \vspace{.3cm}
    \caption{Example of Transition Matrix of RR-Interval Classification for a Random 45-sec/68-beat Sample from the MIT-BIH Dataset}
    \label{wholetmat}
\end{table}

With the transition matrices in Table \ref{wholetmat}, we found the proportions of each transition.

\begin{table}[ht]
    \centering
    \begin{tabular}{llccc}
        & && \textbf{From}&\\
        \toprule
         && \textbf{S}hort & \textbf{N}ormal & \textbf{L}ong \\
         \midrule
         &\textbf{Short} &  0.118 & 0.088 & 0.103 \\
         \textbf{To}&\textbf{Normal} & 0.132 & 0.206 & 0.088 \\
         &\textbf{Long} & 0.059 & 0.147 & 0.059 \\
         \bottomrule
    \end{tabular}
    \vspace{.3cm}
    \caption{Example of Transition Matrix of RR-Interval Classification Proportions for 45-sec/68-beat Sample in Table \ref{wholetmat}}
    \label{tmatprops}
\end{table}

We then unlisted the transmission matrices exemplified in Figure \ref{tmatprops} into the form: [Short-to-Short, Short-to-Normal, Short-to-Long, Normal-to-Short, Normal-to-Normal, Normal-to-Long, Long-to-Short, Long-to-Normal, Long-to-Long]. We also utilized Young et al's $RR_{100}$ algorithm shown in Equation \ref{countit}, and \ref{countclass} and Ghodrati et al's $RR_{var}$ from Equation \ref{rrvar}. We modified both Ghodrati and Young's measurements to better reflect the data.

Ghodrati et al's $RR_{var}$ had a running mean defined by Equation \ref{barrr}, which we changed to reflect Equation \ref{moddyrmean} for consistency. Young et al's $RR_{100}$ algorithm had to be altered to reflect the sample. Because we used the algorithm as a measurement instead of a classification technique we altered the count to become a proportion for different sample sizes. In a preliminary study we tested model performances with Young's 100ms threshold incremented by 10ms. Doing this we found that 200ms was a better threshold for the MIT-BIH and 2017 Challenge datasets. This new measurement was defined as $RR_{200}$:

            \begin{align}
            \label{newcountit}
            \begin{split}
                RR_i - RR_{i-1} > 200ms \rightarrow count = count+1, \\
                RR_i - RR_{i-1} < 200ms \rightarrow count = count-1,
            \end{split}
            \end{align}
            \begin{equation}
               RR_{200} = \frac{count}{\Sigma RR}.
               \label{rr200}
            \end{equation}

We then extracted the length of each subset in beats and seconds, the start and end sample for each subset, and the majority annotation. We did not consider all four rhythm categories within the MIT-BIH dataset for this analysis, instead we dichotomized each 45-second sample as either AFib or not-AFib. This was done to better identify patterns of just Atrial Fibrillation heart-beats and was also deemed by Moody and Mark in 1983 not to significantly affect results. We also did not consider segments that include both AFib and not-AFib for this study because for this analysis we focus on classifying entire segments as AFib or not-AFib. We removed the segments with multiple rhythms for future analysis. With the extracted features we developed a new dataset of just the extracted sample features for each subject. For the 2017 Challenge data we considered both binary and multi-class classification.

\section{Analysis}

\subsection{RR-Intervals and Running Mean}

We first looked at the distribution of the RR-Intervals and running means for each subject using the data from the completed merged datasets. Using plots of the RR-Intervals over the recording period and running means over the recording period we then highlighted all annotations of identified Atrial Fibrillation.

\begin{figure}[H]
    \centering
    \includegraphics[width = .45\textwidth, keepaspectratio]{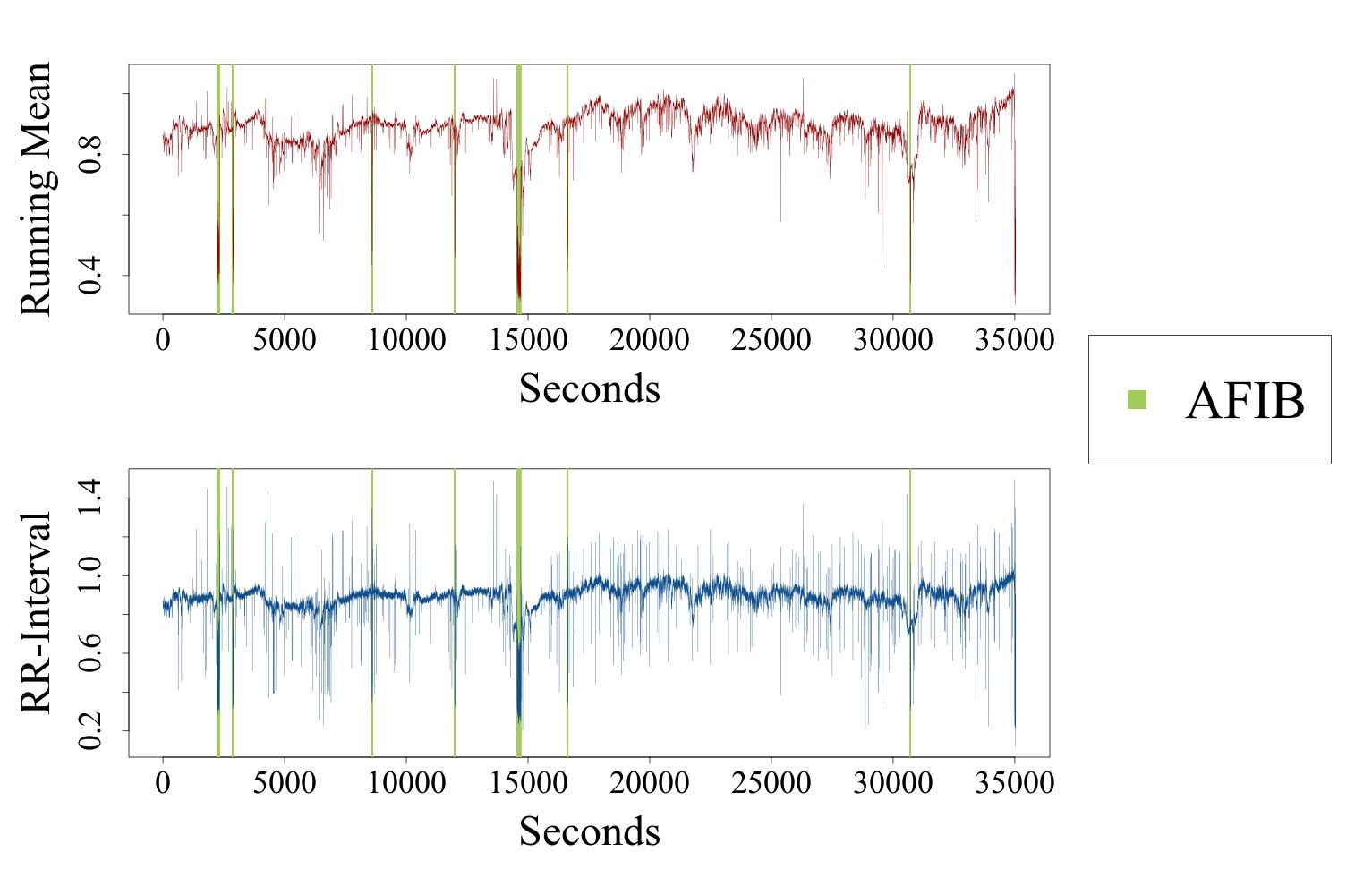}
    \caption{Plot of Running Mean (top) and RR-Intervals (bottom) for Subject 04048 from MIT-BIH Dataset, Highlighted in Green when Subject is in AFib}
    \label{merge04048}
\end{figure}

\begin{figure}[H]
    \centering
    \includegraphics[width =.45\textwidth, keepaspectratio]{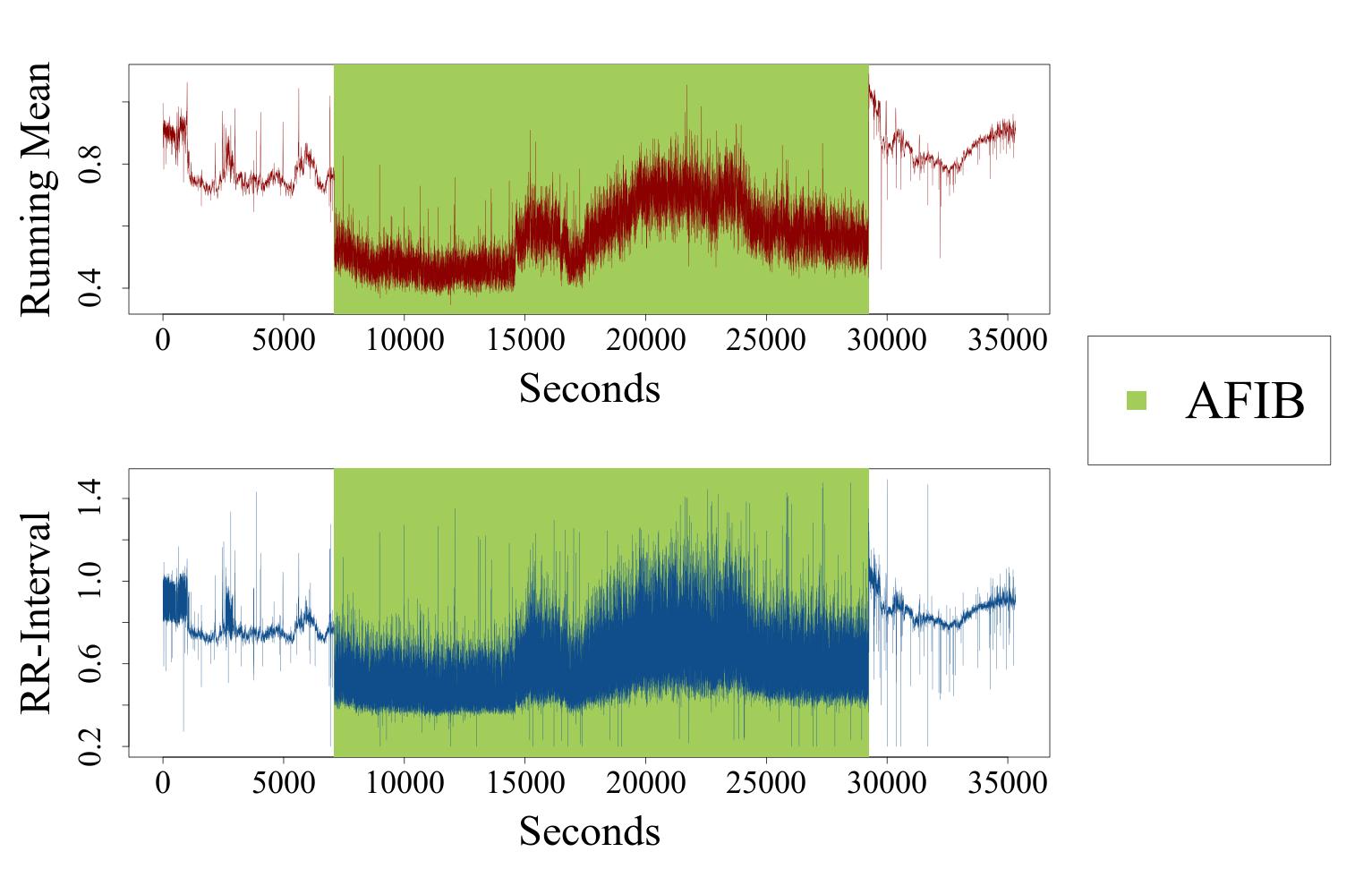}
    \caption{Plot of Running Mean (top) and RR-Intervals (bottom) for Subject 07879 from MIT-BIH Dataset, Highlighted in Green when Subject is in AFib}
    \label{merge07879}
\end{figure}

\begin{table}[ht]
    \centering
    \begin{tabular}{lcc}
    
        & \ \ \ \ \ \ \ \ \ \ \ \ \ \ \ \ \ \ \ \ \ \ \ \ \ \textbf{Subject}\\
    \toprule
         &  \textbf{04048} & \textbf{07879} \\
         \midrule
        \textbf{RR-Int} & 0.009 & 0.032 \\
        \textbf{Running Mean} & 0.007 & 0.022\\
        \bottomrule
    \end{tabular}
    \vspace{.2cm}
    \caption{Variance of RR-Intervals and Running Mean in Subjects 04048 and 07879}
    \label{var1}
\end{table}

Table \ref{var1} shows the variances of both the RR-Intervals and the Running mean over the entire ECG recording for each subject. Even including the sections without AFib, Subject 07879 shows a much higher variance than Subject 04048. This signified to us that looking deeper into the transition proportions we subset previously might also show differences in AFib and not-AFib sections.

\subsection{Transition Matrices Proportions}

\begin{figure*}[h]
    \centering
    \includegraphics[width = .8\textwidth, keepaspectratio]{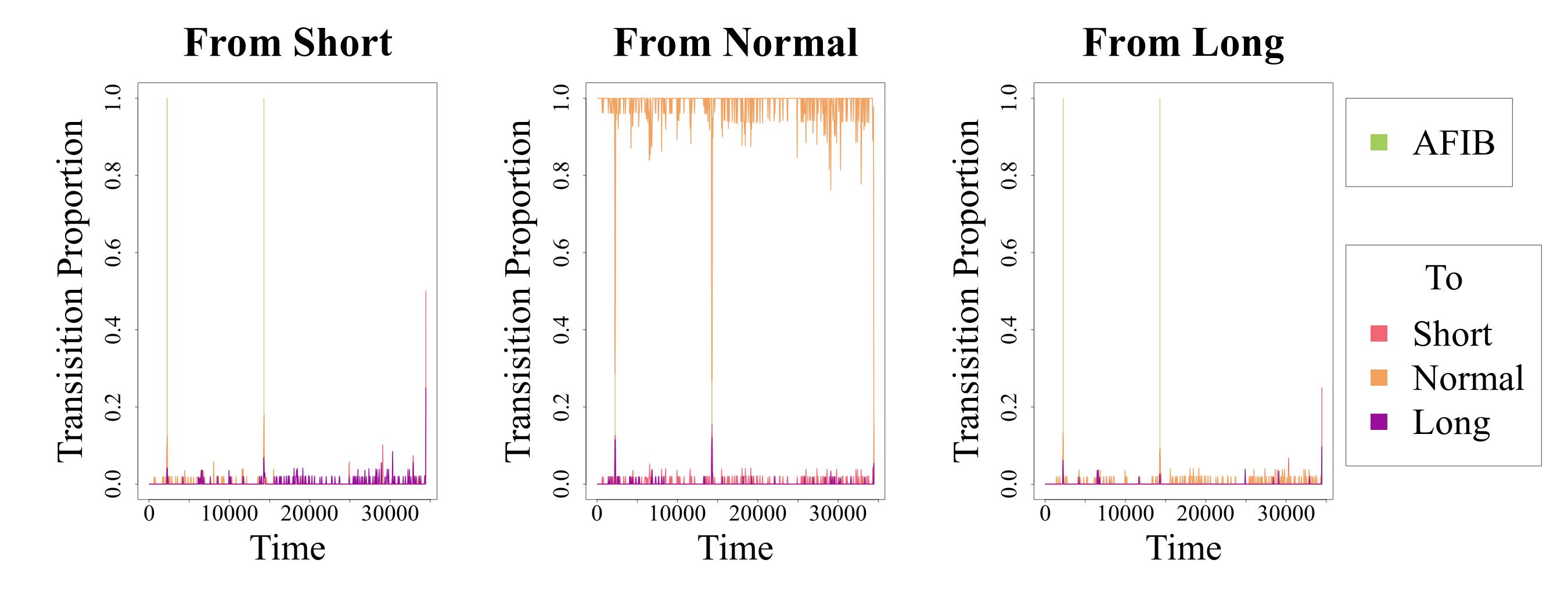}
    \caption{Transition Proportions of each 45-second Subset of Subject 04048 From MIT-BIH Data over Entire Recording Period}
    \label{timetrans04048}
\end{figure*}

\begin{figure*}[h]
    \centering
    \includegraphics[width = .8\textwidth, keepaspectratio]{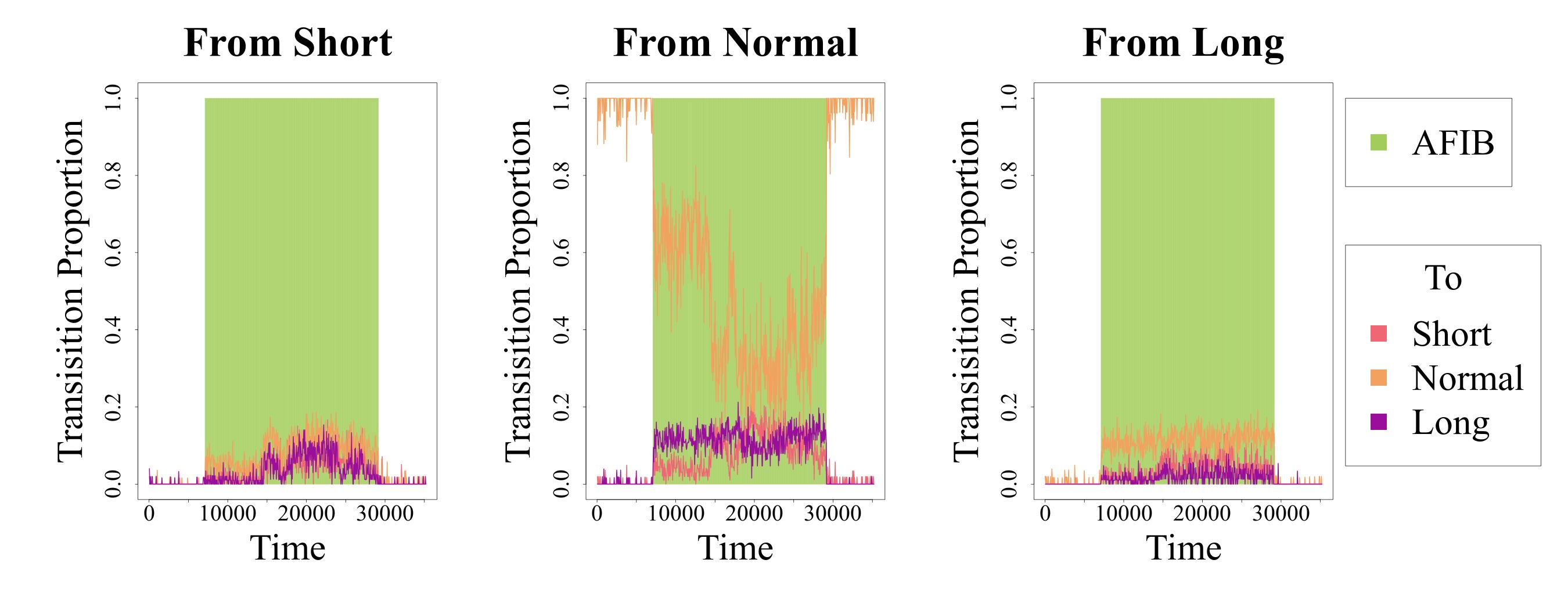}
    \caption{Transition Proportions of each 45-second Subset of Subject 07879 From MIT-BIH Data over Entire Recording Period}
    \label{timetrans07879}
\end{figure*}

Once we identified some patterns of subjects in AFib also having high variance among RR-Intervals and Running Mean, we looked closer at the transition proportions created. We plotted the density of each transition as well as the proportions over the whole recording period. 

Above, Figures \ref{timetrans04048} and \ref{timetrans07879} give the transition proportions extracted over the entire recording period. Where Atrial Fibrillation occurs there is a rise in all lines in `From Short' and `From Long' in Figure \ref{timetrans07879}. In `From Normal,' in Figure \ref{timetrans07879}, the Normal-to-Normal (N-N) proportions begin to converge with N-S and N-L proportions, while in Figure \ref{timetrans04048} and all not-AFib portions of \ref{timetrans07879} N-N is concentrated near 1 and all other transition proportions are very low. This distinct pattern was seen in preliminary studies of all 23 Subjects, details can be found at our website: https://github.com/025georgialynny/afib.

Identifying this pattern in a model does cause some issues because to continuously read in and store transition proportions to develop plots similar to those above it would require a significant amount of memory and time to identify AFib. Because of this problem, we took a deeper investigation into the transition proportions to develop a better understanding of the differences between individuals with low AFib (Subject 04048) and those with a higher amount of AFib (Subject 07879).

\begin{figure*}[h]
    \centering
    \includegraphics[width = .8\textwidth, keepaspectratio]{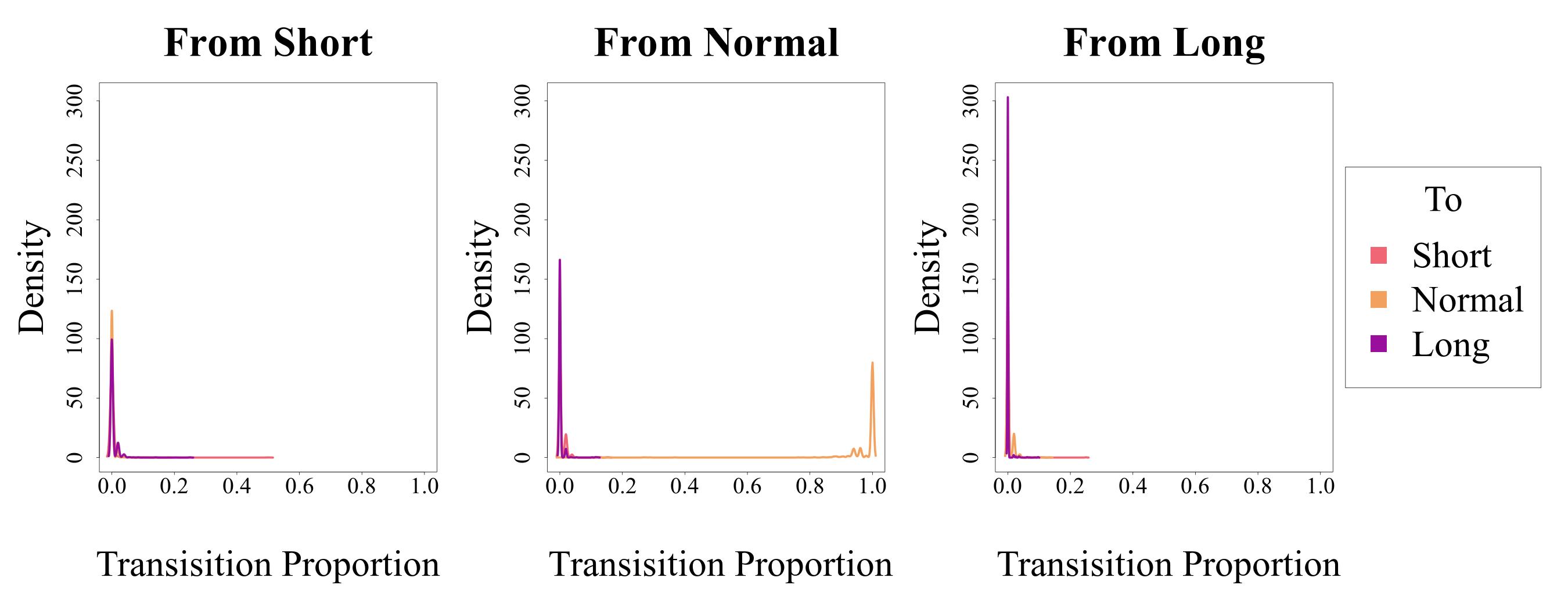}
    \caption{Density of Transition Proportions of the 45-second Subsets of Subject 04048 From the MIT-BIH Dataset over Entire Recording Period}
    \label{densetrans04048}
\end{figure*}

\begin{figure*}[h]
    \centering
    \includegraphics[width = .8\textwidth, keepaspectratio]{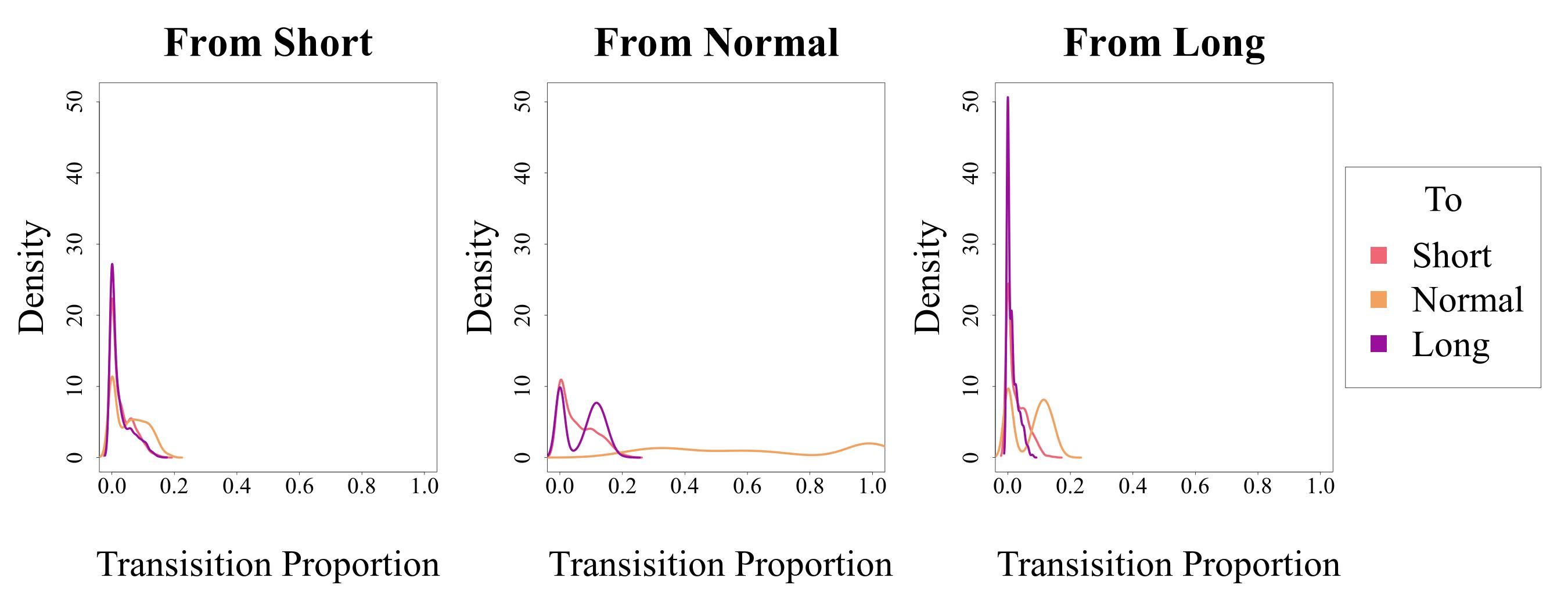}
    \caption{Density of Transition Proportions of the 45-second Subsets of Subject 07879 From the MIT-BIH Dataset over Entire Recording Period}
    \label{densetrans07879}
\end{figure*}

Just looking at the `From Normal' plots in Figures \ref{densetrans04048} and \ref{densetrans07879} there is a clear difference in the spread of the proportions' density. In Figure \ref{densetrans04048}, the data is Bernoulli distributed with binary peaks while Figure \ref{densetrans07879} shows a much more spread distribution of density of proportions. 

\newpage

\section{Classification Models}
\subsection{Model and Feature Selection}

Once the data was subset and features had been extracted we used the MIT-BIH dataset to find the best model to use. This was done because the MIT-BIH dataset is much cleaner giving us a better idea of model performance for AFib-related ECG data. We built our models with MIT-BIH data and then tested with a sample of 300 subjects from the 2017 Challenge data to test the robustness of each model. To do this we dichotomized each sample in the 2017 data as AFib or not AFib to account for the difference in rhythm categories used by the datasets. We compared Logistic Regression (LogReg), Linear Discriminant Analysis (LDA), Quadratic Discriminant Analysis (QDA), Boosting, Support Vector Machines (SVM) with a Radial Kernel (after preliminary analysis of multiple kernels), and Random Forest (RF) as models to classify AFib.

    \begin{table}
    \centering
    \begin{tabular}{l ccc}
    \toprule
         & \textbf{A}ccuracy & \textbf{S}pecificity & \textbf{S}ensitivity \\
         \midrule
        \textbf{LogReg}& 85.619\% & 90.873\% & 57.447\%\\
        \textbf{LDA} &81.605\% & 93.304\% & 46.667\% \\
        \textbf{QDA}&  76.254\% & 82.014\% & 0.000\% \\
        \textbf{Boosting} & 82.274\% & 88.583\% & 53.333\% \\
        \textbf{SVM} & \textbf{87.291\%} & \textbf{88.364\%} & \textbf{75.000\%} \\
        \textbf{RF} & 78.930\% & 95.146\% & 56.989\% \\
        \bottomrule
    \end{tabular}
    \vspace{.2cm}
    \caption{Comparison of Models using MIT-BIH Data to Train and 2017 Challenge to Test}
    \label{3covmodcomp}
\end{table}

Support Vector Machines, with a radial kernel, outperformed the other models, especially just looking at the Sensitivity of the models. Because Atrial Fibrillation is a serious health disorder we are willing to sacrifice sensitivity for specificity to increase predictions of AFib at a reasonable rate. With this in mind we continued with SVM with a radial kernel as our preferred model.

\begin{table*}
\centering
    \begin{tabular}{ll}
    \toprule
        \textbf{L}abel & \textbf{K}ey \\
        \midrule
        \textbf{Model 1} & 9 Transition Proportions, $RR_{var}$, and $RR_{200}$  \\
        \textbf{Model 2} & Normal-Normal, $RR_{var}$, and $RR_{200}$ \\
        \textbf{Model 3} & Normal-Long, $RR_{var}$, and $RR_{200}$ \\
        \textbf{Confusion Matrix Columns} & Reference Classifications\\
        \textbf{Confusion Matrix Rows} & Predicted Classifications \\
        \bottomrule
    \end{tabular}
    \caption{Key for Table \ref{bigtab}}
    \label{tab:key}
\end{table*}

\begin{table*}[ht]
    \centering
    \begin{tabular}{lcccc}
    \toprule
    & \textbf{C}onfusion \textbf{M}atrix &\textbf{ A}ccuracy & \textbf{S}pecificity & \textbf{S}ensitivity \\
    \midrule
    \textbf{Model 1} & \begin{tabular}{lcc}
        & \textbf{N} & \textbf{AF}  \\
         \midrule
        \textbf{N} & 10,386 & 144 \\
        \textbf{AF} & 217 & 6,861 \\
    \end{tabular} & 97.948\% & 97.965\% & 97.944\% \\
    \midrule
        \textbf{Model 2} & \begin{tabular}{lcc}
         & \textbf{N} & \textbf{AF} \\
         \midrule
        \textbf{N} & 9,652 & 300 \\
        \textbf{AF} & 951 & 6,705 \\
    \end{tabular} & 92.895\% & 91.031\% & 95.717\% \\
    \midrule
    \textbf{Model 3} & \begin{tabular}{lcc}
         & \textbf{N} & \textbf{AF} \\
         \midrule
        \textbf{N} & 10,285 & 175 \\
        \textbf{AF} & 318 & 6,830 \\
    \end{tabular} & 97.200\% & 97.001\% & 97.502\% \\
    \bottomrule 
    \end{tabular}
    \vspace{.2cm}
    \caption{Performance of Models with Different Prediction Variables}
    \label{bigtab}
\end{table*}

\begin{table*}[!htbp]
    \centering
    \begin{tabular}{cccc}
    \toprule
    \textbf{C}onfusion \textbf{M}atrix &\textbf{A}ccuracy & \textbf{S}pecificity & \textbf{S}ensitivity \\
    \midrule
    \begin{tabular}{lcc}
        & \textbf{N} & \textbf{AF}  \\
         \midrule
        \textbf{N} & 10,340 & 263 \\
        \textbf{AF} & 195 & 6,810 \\
    \end{tabular} & 97.399\% & 98.149\% & 96.282\% \\
        \bottomrule
    \end{tabular}
    \caption{LOO-CV on MIT-BIH Data with Normal-to-Long Transition Proportion, $RR_{200}$, and $RR_{var}$ Performance}
    \label{tab:rtolsvmloocv}
    \vspace{-.5cm}
\end{table*}

A problem with all models is the collinearity of our features, since all of the transition proportions summed to one. For further model testing in the future the collinearity will need to be reduced for model validation. To reduce the collinearity without the computational intensity of dimension reduction we compared the results using all transition proportions with $RR_{200}$ and $RR_{var}$ with just Normal-to-Long or Normal-to-Normal with $RR_{200}$ and $RR_{var}$. 

With just Normal-to-Long paired with $RR_{200}$ and $RR_{var}$ as predictor variables the SVM model performs very close results to all Model 1 This is very preferable because it prevents collinearity for comparison of SVM with other models and decreases the computational intensity of the model for future implementation.

Leave-One-Out Cross-Validation (LOO-CV) performed very similarly with Normal-to-Long, $RR_{200}$, and $RR_{var}$ in Table \ref{tab:rtolsvmloocv}. LOO-CV performance is important in this analysis because in a real-time application each 45-second segment will be classified individually, instead of in a large group. Because LOO-CV has a comparable performance to the other models we'd run we are more confident in these features' ability to represent/classify Atrial Fibrillation. 

\subsection{Exploratory Analysis}

\begin{table*} 
\centering
    \begin{tabular}{p{.47\textwidth}p{.47\textwidth}}
    \includegraphics[width = .45\textwidth, keepaspectratio]{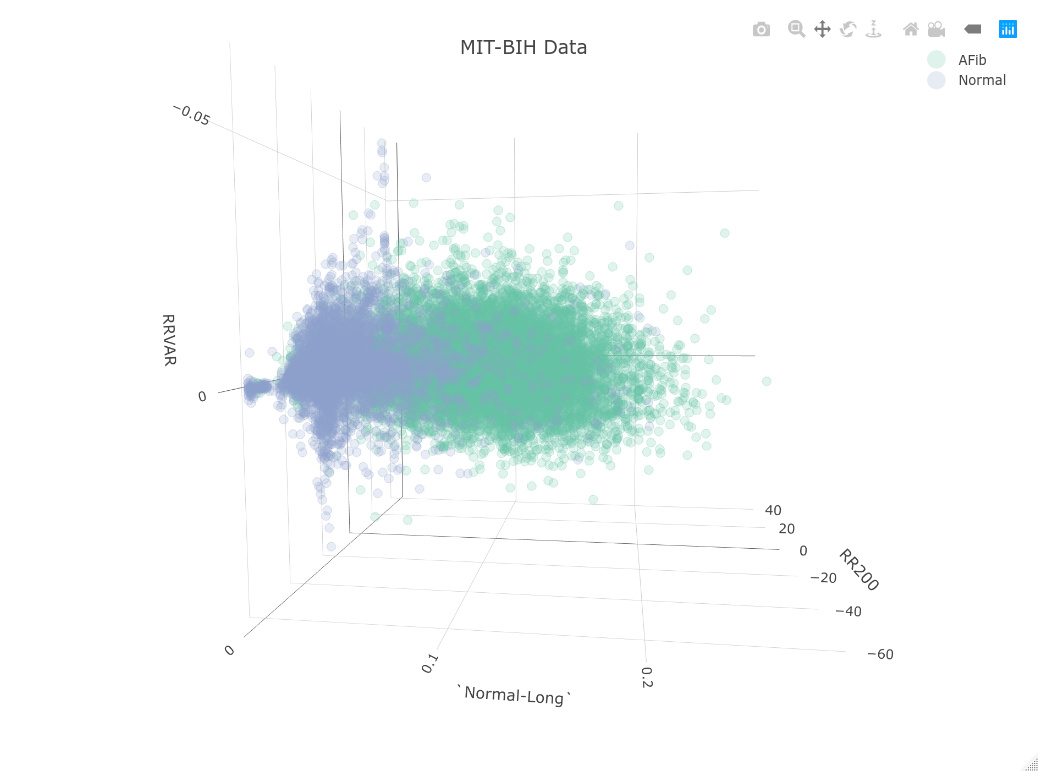} &
    \includegraphics[width = .45\textwidth, keepaspectratio]{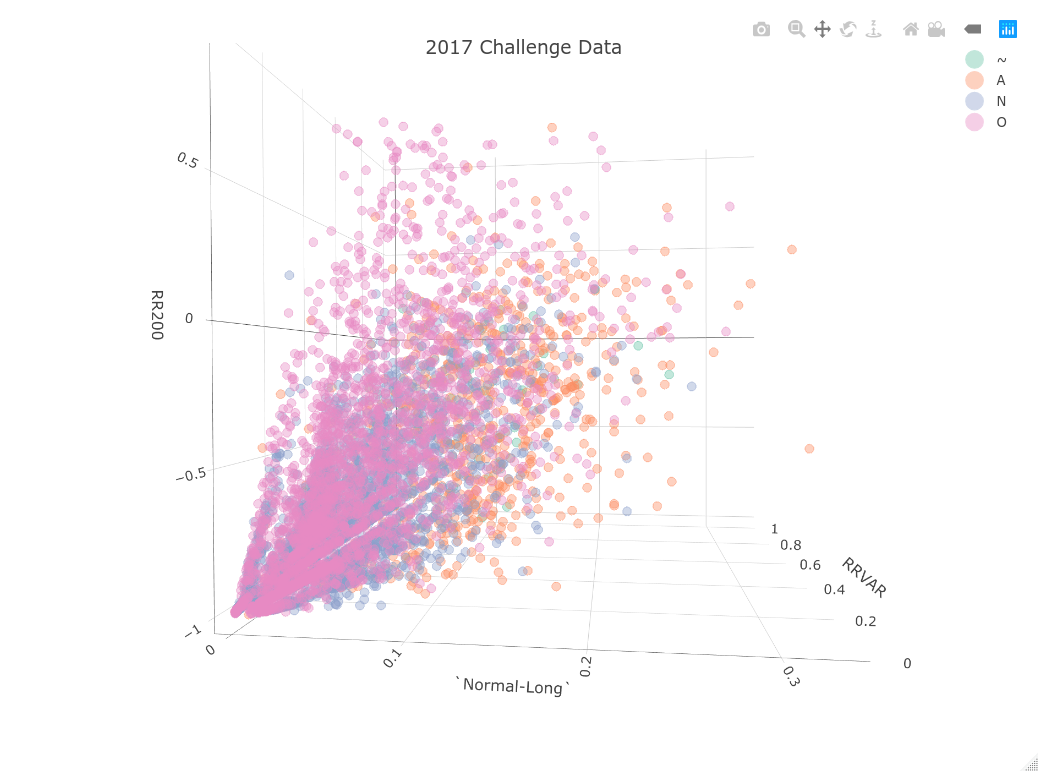}\\
    a). Dichotomized MIT-BIH Data with Normal-to-Long Transition, $RR_{200}$, and $RR_{var}$ &
    b). Plot 1 of 2017 Challenge Data with Normal-to-Long Transition, $RR_{100}$, and $RR_{var}$ \\
        
    \includegraphics[width = .45\textwidth, keepaspectratio]{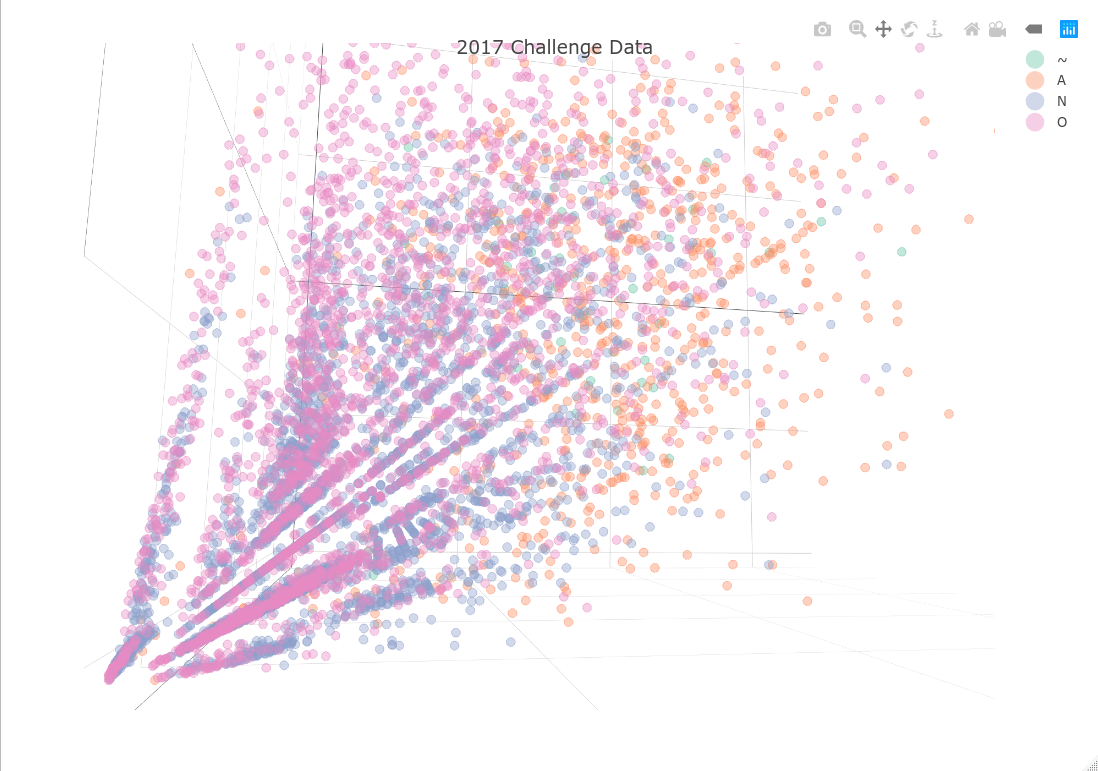}&
    \includegraphics[width = .45\textwidth, keepaspectratio]{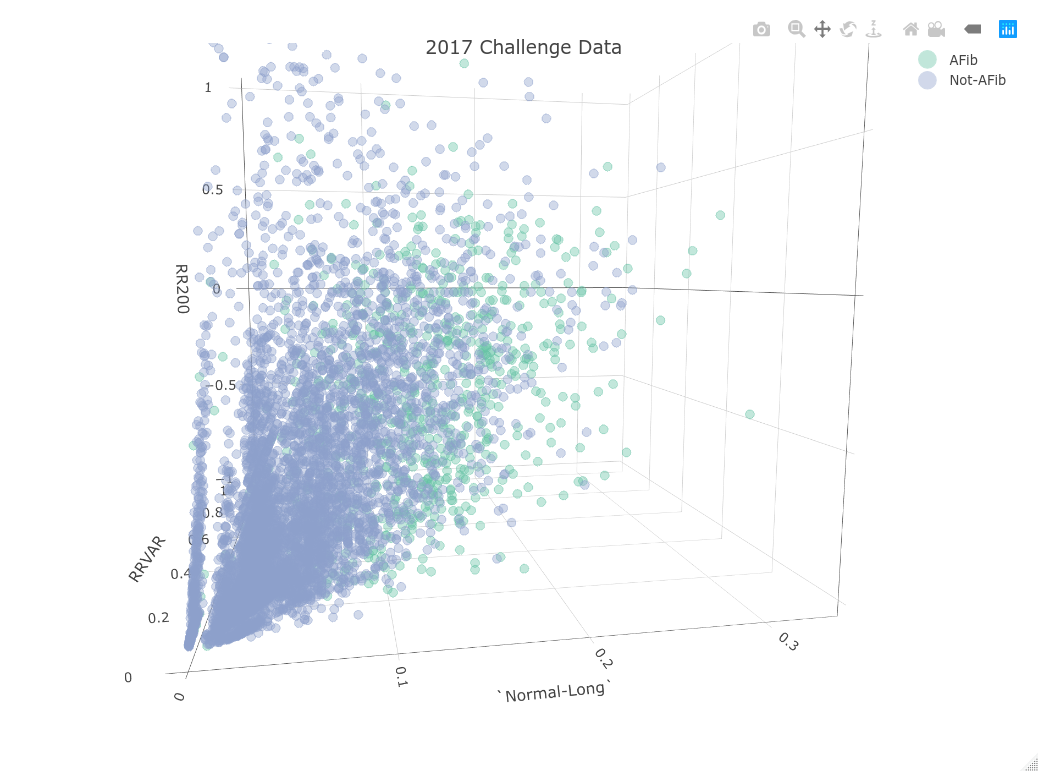} \\
    
    c). Plot 2 of 2017 Challenge Data with Normal-to-Long Transition, $RR_{100}$, and $RR_{var}$ &
    d). Dichotomized 2017 Challenge Data with Normal-to-Long Transition, $RR_{100}$, and $RR_{var}$ \\
    
    \end{tabular}
    \caption{Exploratory Figures from MIT-BIH Dataset and 2017 Challenge Dataset}
    \label{dataplots}
\end{table*}

With three covariates the data was reduced to a visible dimension, allowing us to visually analyze the relationship between the heart-beat rhythms and the extracted features/covariates.

With the MIT-BIH data good separation between Atrial Fibrillation and not-Atrial Fibrillation occurs with just Normal-to-Long, $RR_{100}$, and $RR_{var}$ as seen in Figure \ref{dataplots}-a. Although, transitioning to the 2017 Challenge data does not give the same clear separation with Normal-to-Long, $RR_{var}$, and $RR_{200}$ as the only predictor variables.

In Figure \ref{dataplots}-b, with all four classifications, the 2017 Challenge data shows little to no separation of the data.

When zoomed in further in Figure \ref{dataplots}-c it is observed the Normal and Other heart rhythms follow near identical patterns within the data while Atrial Fibrillation is slightly away from it. This implies that the heart-rhythms categorized as other by Physionet are not identifiable with RR-Intervals, or the features we extracted for this analysis. 

In Figure \ref{dataplots}-d, with Other and Normal considered the same rhythm, there is more of difference between the rhythms. While the separation is not as pronounced as with the MIT-BIH database, finding the right amount of Support Vectors within our Radial SVM may result in good results.

\subsection{Transfer Learning Using Support Vector Machines}

To analyze transfer learning with the SVM model we built, two new models were developed. They were a balanced model developed with 5,000 AFib samples and not-AFib samples from the MIT-BIH Dataset, as well as a weighted model developed with all samples from the MIT-BIH Dataset. These two models were then tested on the 2017 Challege Data with each reading once again dichotomized as AFib or not-AFib for consistency with the MIT-BIH data. 

The first model was trained with the balanced dataset of 5,000 AFib and 5,000 not-AFib from the MIT-BIH data set. It was then tested on the 2017 Challenge data set. In our preliminary studies we tested different parameters and found cost = 10 and $\gamma = \frac{1}{18}$ had the best performance. This model was compared against SVM with the entire dataset of 17,613 MIT-BIH samples with uneven amounts of not-AFib and AFib samples. The class weights which were used for this are given by 

\begin{equation}
    [AFib, \text{\textit{not-AFib}}] = [\frac{1 - \Sigma AFib}{17,613}, \frac{1 - \Sigma \text{\textit{not-AFib}}}{17,613}]
    \label{weightsSVM1},
\end{equation}

where $\Sigma \text{\textit{not-AFib}}$ and $\Sigma AFib$ are the number of not-AFib or AFib samples, respectively, within the training set.

\begin{table}[ht]
    \centering
    \begin{tabular}{lccc}
    \toprule
         & \textbf{A}ccuracy & \textbf{S}ensitivity & \textbf{S}pecificity  \\
         \midrule
        \textbf{Balanced SVM} & 84.28\% & 92.44\% & 53.46\% \\
        \textbf{Weighted SVM} & 83.95\% & 93.13\% & 51.52\% \\
        \bottomrule
    \end{tabular}
    \vspace{.2cm}
    \caption{SVM Testing Results}
    \label{SVM1results}
\end{table}

As seen in Table \ref{SVM1results}, the balanced SVM model performed slightly better in sensitivity than the weighted model. To address this models are trained and tested on the 2017 Challenge data to filter out any cross database model problems.

To classify the 2017 Challenge data a Support Vector Machine Pipeline, or multiple Support Vecor Machines, described below, was required to account for each classification provided with the data.

\begin{enumerate}
    \item Train SVM model with Noisy classifications against all alternative classifications
    \item Remove what is predicted as Noisy from the Testing Set and what is Noisy in the Training Set
    \item Train SVM model with AFib classifications against all remaining classifications
    \item Remove what is predicted as AFib from the Testing Set and what is Noisy in the Training Set
    \item Train SVM model with Other classifications vs all remaining classifications
    \item Remove what is predicted as Other from the Testing Set
    \item Classify left-over data as Normal
\end{enumerate}

With this ``pipeline" we performed 5-fold Cross-Validation on the 2017 Challenge Dataset.

    \begin{table}[ht]
    \centering
    \begin{tabular}{l ccc}
    \toprule
         & \textbf{A}ccuracy & \textbf{S}pecificity & \textbf{S}ensitivity \\
         \midrule
        \textbf{Noise}& 99.479\% & 99.526\% & 25.000\%\\
        \textbf{AFib} &92.115\% & 94.213\% & 58.964\% \\
        \textbf{Other}&  75.177\% & 78.885\% & 61.061\% \\
        \bottomrule
    \end{tabular}
    \vspace{.2cm}
    \caption{Results of 5-Fold Cross Validation of Pipeline SVM Model using 2017 Challenge Data}
    \label{2017piperez}
    \end{table}
    
Table \ref{2017piperez} shows that using the ``pipeline" for multiclass classification did not have high performance, with similar results to training with the MIT-BIH data set and testing with the 2017 Challenge data set. While the Specificities are moderate and the Accuracies are not extremely low, the Sensitivities can be improved for detection of a life threatening disease such as Atrial Fibrillation. While this approach does have potential for Atrial Fibrillation diagnosis, the use of all four classifications may be causing inaccuracies in the SVM models.

Based on the plots of the 2017 Challenge Dataset we ran binary classification SVM on the 2017 Challenge dataset. The results were much higher, supporting our previous hypothesis that RR-Intervals and features based on RR-Interval variation are not sufficient for classifying all Other rhythms.

\begin{table}[ht]
    \centering
    \begin{tabular}{cccc}
    \toprule
           \textbf{C}onfusion \textbf{M}atrix &\textbf{ A}cc & \textbf{S}pec & \textbf{S}ens \\
           \midrule
         \begin{tabular}{lcc}
                & \textbf{N} & \textbf{AF} \\
                \midrule
                \textbf{N} & 6,488 & 142 \\
                \textbf{AF} & 1,257 & 626 \\
           \end{tabular} & 0.836 & 0.838 & 0.815 \\
           \bottomrule
           \end{tabular}
           \vspace{.2cm}
    \caption{Binary SVM Results of 2017 Challenge Dataset}
    \label{tab:binary2017chall}
\end{table}

\section{Conclusions}

When looking at the data in the MIT-BIH Database and 2017 Challenge Dataset we were given the RR-Intervals and Annotations of the heart-beat Rhythm Types. Once these were extracted we could identify features to identify patterns to understand the data. We extracted features of transitions based on the running mean and classifying the heart-beats as Normal, Short, or Long compared to the running mean. Doing this we found that in AFib the transitions are much less binary and tend to converge when transitioning from the Normal heart-beats to other classifiers. With this understanding of the data we continued to extract features such as $RR_{var}$ and $RR_{200}$.

With these features extracted we identified Support Vector Machine with a radial kernel as a promising model to classify AFib. When tested with the MIT-BIH database this method worked very well and led us to decrease our covariates to just Normal-To-Long Transition Proportion, $RR_{var}$, and $RR_{200}$. The plot of the MIT-BIH data further supported this decision showing clear separation of AFib and not-AFib rhythms. When switching databases we did run into some problems.

Plotted, the 2017 Challenge dataset did not show any clear separation with just three covariates and when tested with SVM trained on the MIT-BIH database it did not perform in a promising manor. To account for any cross-database inconsistencies we tested and trained an SVM model on the 2017 Challenge data in a ``pipeline" approach to account for all four classifications supplied with the data. This approach did increase our statistics, but sensitivity was not raised to a desirable level for our goal.

With binary classification the sensitivity was increased, implying that RR-Intervals are not sufficient data for all heart-beat rhythms. Although, features based on RR-Intervals are well suited for Atrial Fibrillation classification. 
\section{Future Work}

Classification of the MIT-BIH data worked very well with just three predictor variables and a Radial SVM model. Although, this dataset is very oversampled and highly processed/cleaned. Cross validation with the 2017 Challenge data resulted in lower performance when four heart-rhythms were classified, but increased with binary classification. This implies that while RR-Interval-based features may be suited to classify Atrial Fibrillation, they are not suited to classify all heart rhythms. For application in Photoplethysmographs this hypothesis needs to be cross validated with PPG datasets. While cleaning of the 2017 Challenge ECG data did improve performance, the same processing techniques may not work with PPG data.

In the future, this research will be continued to look closer into processing an classification with PPG data, as well as application into a real time model that can filter out noise and extract features continuously. 
\section*{Acknowledgements}
\textit{This work was supported by the National Science Foundation under DMS Grant Number 1659288. I would like to thank Dr. Cuixian Chen and Dr. Yishi Wang for this opportunity, their support in learning the methods necessary for this research, and pushing me to learn more. Thank you to my colleagues Christian Thompson, Jericho Lawson, and Drew Johnston for their input and support through this process.}
{\small
\bibliographystyle{ieee}
\bibliography{egbib}

\begin{thebibliography}{1}\itemsep=-1pt

\bibitem{clifford2017af}
G.~D. Clifford, C.~Liu, B.~Moody, H.~L. Li-wei, I.~Silva, Q.~Li, A.~Johnson,
  and R.~G. Mark.
\newblock Af classification from a short single lead ecg recording: the
  physionet/computing in cardiology challenge 2017.
\newblock In {\em 2017 Computing in Cardiology (CinC)}, pages 1--4. IEEE, 2017.

\bibitem{wfdb}
T.~Edition and G.~B. Moody.
\newblock Wfdb applications guide.

\bibitem{moody1983}
R.~M. GB~Moody.
\newblock A new method for detecting atrial fibrillation using r-r intervals.
\newblock {\em Computers in Cardiology}, 10:227--230, 1983.

\bibitem{ghodrati2008statistical}
A.~Ghodrati and S.~Marinello.
\newblock Statistical analysis of rr interval irregularities for detection of
  atrial fibrillation.
\newblock In {\em 2008 Computers in Cardiology}, pages 1057--1060. IEEE, 2008.

\bibitem{goldberger2000physiobank}
A.~L. Goldberger, L.~A. Amaral, L.~Glass, J.~M. Hausdorff, P.~C. Ivanov, R.~G.
  Mark, J.~E. Mietus, G.~B. Moody, C.-K. Peng, and H.~E. Stanley.
\newblock Physiobank, physiotoolkit, and physionet: components of a new
  research resource for complex physiologic signals.
\newblock {\em Circulation}, 101(23):e215--e220, 2000.

\bibitem{lake2010accurate}
D.~E. Lake and J.~R. Moorman.
\newblock Accurate estimation of entropy in very short physiological time
  series: the problem of atrial fibrillation detection in implanted ventricular
  devices.
\newblock {\em American Journal of Physiology-Heart and Circulatory
  Physiology}, 300(1):H319--H325, 2010.

\bibitem{rubio}
B.~Larburu~Rubio.
\newblock Estudio comparativo de algoritmos para la detecci{\'o}n de la
  fibrilaci{\'o}n auricular.
\newblock 2011.

\bibitem{nihbli}
NIHBLI.
\newblock Atrial fibrillation.

\bibitem{young1999comparative}
B.~Young, D.~Brodnick, and R.~Spaulding.
\newblock A comparative study of a hidden markov model detector for atrial
  fibrillation.
\newblock In {\em Neural Networks for Signal Processing IX: Proceedings of the
  1999 IEEE Signal Processing Society Workshop (Cat. No. 98TH8468)}, pages
  468--476. IEEE, 1999.

\end{thebibliography}
}

\end{document}